\documentclass[pra,twocolumn,superscriptaddress]{revtex4}

\usepackage{amsmath,amsfonts,amssymb,slashed}
\usepackage[english]{babel}
\usepackage{graphicx}
\usepackage{color}
\usepackage{mathrsfs}
\usepackage{subfigure}
\usepackage{psfrag}

\newcommand{\etilde}{\tilde{\mathcal{E}}}

\begin{document}

\title{Probing the topological properties of the Jackiw-Rebbi model with light}

\author{Dimitris G. Angelakis}
\email{dimitris.angelakis@gmail.com}
\affiliation{School of Electronic and Computer Engineering, Technical University of Crete, Chania, Crete, Greece, 73100}
\affiliation{Centre for Quantum Technologies, National University of Singapore, 2 Science Drive 3, Singapore 117542.}

\author{Priyam Das}
\affiliation{Centre for Quantum Technologies, National University of Singapore, 2 Science Drive 3, Singapore 117542.}

\author{Changsuk Noh}
\email{cqtncs@nus.edu.sg}
\affiliation{Centre for Quantum Technologies, National University of Singapore, 2 Science Drive 3, Singapore 117542.}

\begin{abstract}
The Jackiw-Rebbi model describes a one-dimensional Dirac particle coupled to a soliton field and can be equivalently thought of as the model describing a Dirac particle under a Lorentz scalar potential. Neglecting the dynamics of the soliton field, a kink in the background soliton profile yields a topologically protected zero-energy mode for the particle, which in turn leads to  charge fractionalization. We show here that the model can be realized in a driven slow-light setup, where photons mimic the Dirac particles and the soliton field can be implemented-and tuned-by adjusting optical parameters such as the atom-photon detuning.  Furthermore, we discuss how the existence of the zero-mode, and its topological stability, can be probed naturally by analyzing the  transmission  spectrum.  We conclude by doing an analysis of the robustness of our approach against  possible experimental errors in engineering the Jackiw-Rebbi Hamiltonian in this optical set up.

\end{abstract}

\maketitle

\section{Introduction}

Quantum simulators first and foremost offer a promising alternative when analytical and numerical methods fail in analyzing models with strong correlations. On the other hand, they can also be used in probing exotic  physical phenomena such as those predicted by relativistic theories. To date, a collection of effects in different fields ranging from condensed matter physics to relativistic quantum theories and material science have been simulated \cite{CiracZoller2012a}, using different platforms such ion traps \cite{RoosBlatt2012},  and cold atoms in optical lattices \cite{BlochNascimbene2012}.

More recently, the ability to controllably manipulate photons and their interactions with atomic systems, resulted in the birth of a new direction in quantum simulations using photons and polaritons to mimick strongly correlated phenomena. Coupled cavity QED arrays (CCAs) were initially considered, where photons trapped in resonators interfaced with two level atoms (real or artificial ones) were shown to be able to reproduce many body dynamics \cite{reviews}. The so-called photon blockade effect was exploited and polariton Mott transitions and effective spin-models were proposed, introducing what is now known as the Jaynes-Cummings-Hubbard model (JCH) \cite{SIP2}. Simultaneously and independently the possibility for strongly correlated polariton dynamics in CCAs with four level atoms and external fields was proposed \cite{SIP1}, followed soon after by the JCH's phase diagram \cite{SIP3,RossiniFazio07} and the Fractional Hall effect \cite{ChoBose08}. More recently one dimensional highly nonlinear waveguides with slow light nonlinearities \cite{Lukinreview,Marangosreview,Bajcsy2009}  have been considered where effects characterizing Tonks gases,  Luttinger liquids physics or even interacting relativistic theories have been shown to be simulable \cite{Chang,Angelakis11,Angelakis}. The possibility to probe out of equilibrium phenomena has also been explored in driven set ups \cite{carusotto2009fermionized,grujic2012angelakis,reviews}

Quantum simulation has also shown great development in bringing the exotic physics of single particle relativistic effects into laboratory. These range from numerous theoretical works studying the Dirac equation and emerging effects such as the trembling motion of the electron (Zitterbewegung) or Klein tunneling to experimental implementations of those  in different platforms covering all three mentioned above. More specifically in ion technologies the Dirac equation  in 1+1  dimensions has been realized \cite{Lamata, Gerritsma, Casanova,Gerritsma11} followed by implementation with photons in waveguide arrays, including the random mass Dirac model \cite{Longhi,Dreisow,Keil}. Seminal proposals for the realization in slow light systems also exist \cite{Juzeliunas,Otterbach,Unanyan,Ruseckas}. Going beyond the Dirac equation, recent works have proposed to simulate the Majorana equation \cite{Casanova1,Noh} and neutrino oscillations \cite{Noh1} in trapped ions.

The purpose of this article is to propose a quantum simulation of  a historically important relativistic model known as the Jackiw-Rebbi (J-R) model \cite{Jackiw} with slow light. The model describes a one-dimensional Dirac particle coupled to a static soliton field and can be equivalently thought of as the model describing a massless Dirac particle under a Lorentz scalar potential. The same model has been studied independently by Su, Schrieffer, and Heeger to describe electron-phonon coupling in polyacetylene \cite{Su}. 
The model is well-known for predicting charge fractionalization\cite{Niemi}, well before fractional quantum Hall effect was discovered, and also for the topological nature of its zero-energy solution and can be thought of as a precursor to topological insulators, a topic that is being hotly pursued right now \cite{TIreview1,TIreview2}. 

There have been proposals to realize the model and observe charge fractionalization in a optical lattices setup \cite{RuostekoskiJavanainen2002,JavanainenRuostekoski2003} and experimental observation of the soliton which follows the model in a fermionic superfluid \cite{YefsahZwierlein2013}. Also  quantum walk and  graphene realizations exist  of topological bound states analogous to that arising in the model 
\cite{KitagawaWhite2012,YouMundry2007,RomanovskyLandman2013}. In this work, we follow a different route and propose a photonic implementation of the model in a slow light polaritonic setup and show that the topological properties can be probed straightforwardly in an optical transmission experiment. We would like to highlight here that the usual experimental difficulties, in realizing the photonic nonlinear interaction in slow light systems for more complex many body simulations, do not exist in this case. Therefore we believe the proposal is a good candidate for directly realizing the J-R model and could allow the efficient probing of its topological properties for the first time. 

\section{Solitons and the Dirac equation: The case of Jackiw-Rebbi model}

The Jackiw-Rebbi model and the related model studied by Su, Schrieffer, and Heeger \cite{Su} share many features similar to those studied in topological insulators and might indeed be classified as an AIII-type chiral topological Dirac insulator under a suitable regularization \cite{Ryu}. Here we describe the Jackiw-Rebbi model and point out the topological properties and similarities to topological insulators.

The system considered by Jackiw and Rebbi consists of a Dirac particle coupled to a scalar field, which acts as a position dependent mass term. The equation of motion for the Dirac spinor reads
\begin{align}
\label{JRmodel}
i\partial_t{\bf \Psi} = \left( \alpha c p_z + \frac{\beta m c^2}{\kappa} \phi(z) \right) {\bf\Psi},
\end{align}
where $\alpha$ and $\beta$ are the Dirac matrices which in this case can be chosen to be proportional to two Pauli matrices. For concreteness, we take them to be $\alpha = -\sigma_z$ and $\beta = \sigma_y$. The real scalar field $\phi(x)$ is assumed to obey the Klein-Gordon equation with the potential energy of the form
\begin{align}
\frac{\lambda^2}{2\kappa^2} \left( \kappa^2 - \phi(z)^2\right)^2.
\end{align}

\begin{figure}[ht]
\begin{center}
\includegraphics[width=4.0cm]{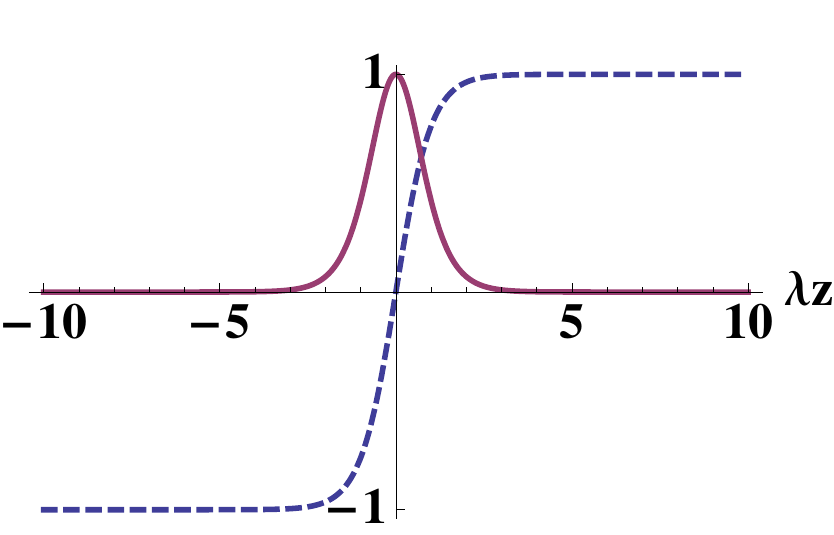}
\end{center}
\caption{Soliton profile, $\phi_s(z)$ (blue dotted), and the zero mode wavefunction, $|\Psi_0(z)|^2$ (red solid), showing the localization scale $\lambda$ of the zero mode. We have set $m=c=\kappa=1$ for convenience.}
\label{zeromode}
\end{figure}

The ground state of the scalar field is degenerate at $\phi(z) = \pm\kappa$ and the degeneracy implies the existence of a soliton that interpolates between $-\kappa$ at $z= -\infty$ and $\kappa$ at $z=\infty$ and corresponding anti-soliton. The soliton localized at $z=0$ is described by 
\begin{align}
\phi_s(z) = \kappa \tanh(\lambda z)
\end{align}
When the Dirac field is coupled to such a soliton, a zero mode (zero energy state) appears which is localized around the soliton. The unnormalized spinor wave function of the zero mode reads
\begin{align}
\label{zeromodewf}
{\bf\Psi}_0(z) &= \exp\left( -\frac{mc}{\kappa}\int_0^z dx \phi_s(x) \right) {\bf\chi} \nonumber \\ &=  \exp\left[ -\frac{mc}{\lambda}\ln(\cosh \lambda z) \right] {\bf\chi},
\end{align}
where $\alpha \beta \bf{\chi}$ $= -i\bf{\chi}$, and is shown in Fig.~\ref{zeromode}. Note that the dynamics of the scalar field is neglected in the above argument, i.e., the scalar field is treated as a constant background field. 

Far away from the kink the particle and hole bands of the Dirac particle have an energy gap $m$, whereas this gap must close at the point $\phi_s(z) = 0$. This resembles the gap closing at the boundary of a topological insulator where the bound surface mode develops. The resulting bound state, called the zero-mode is protected by the topology of the scalar field, whose existence, irrespective of the local profile of the kink, is guaranteed by the so-called index theorem \cite{Callias,Jackiw}. This phenomenon is similar to the emergence of edge modes in the quantum Hall effect \cite{QHE1,QHE2} or topological insulators \cite{TIreview1,TIreview2}, where edge modes appear at the boundary of two topologically different domains.

Another interesting aspect of the model (when second quantization is taken into account) is charge fractionalization, which we briefly describe before we move on to a proposal for a photonic implementation. The ground state of the Dirac field (the vacuum) in the soliton background may or may not include the zero mode. Because of the charge conjugation symmetry of the system, the two degenerate ground states must have opposite charges. Moreover, the charge difference between the two must be 1, as there can be only one electron occupying the state. The result is that the filled or unfilled zero modes must have charges 1/2 and -1/2, respectively. This is confirmed by constructing the formal charge operator in terms of creation and annihilation operators of the eigen-modes.  The vacuum states are the eigenvalues of the charge operators which means that the observed fractional charges are sharp and are not just a trivial realization of a distributed charge. 

\begin{figure}[ht]
\begin{center}
\subfigure[$ $]{
\includegraphics[scale=0.25]{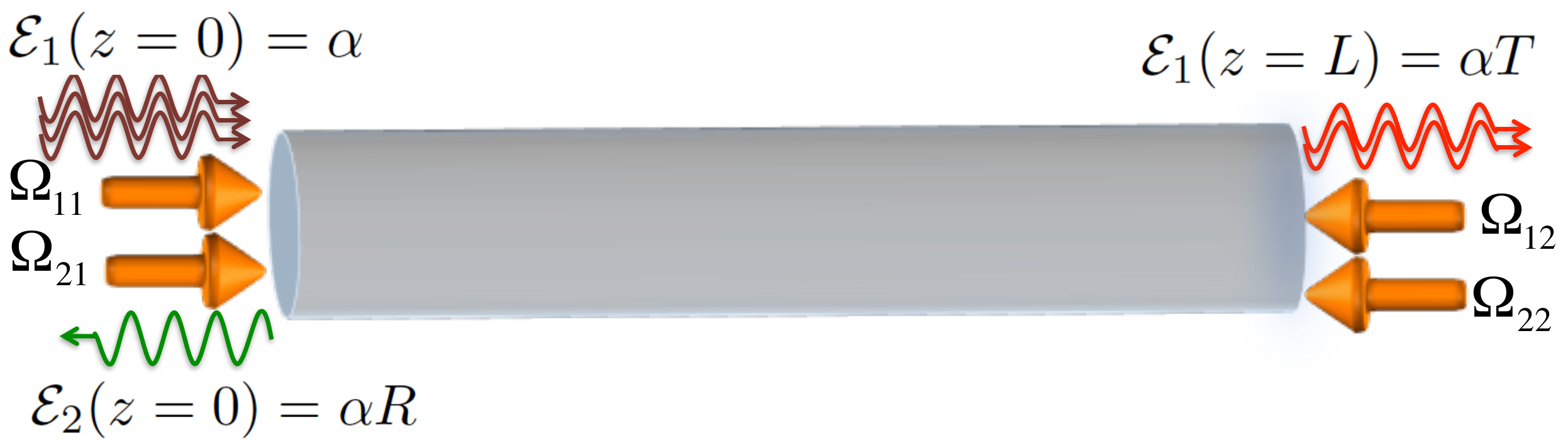}} \\
\subfigure[$ $]{
\includegraphics[scale = 0.25]{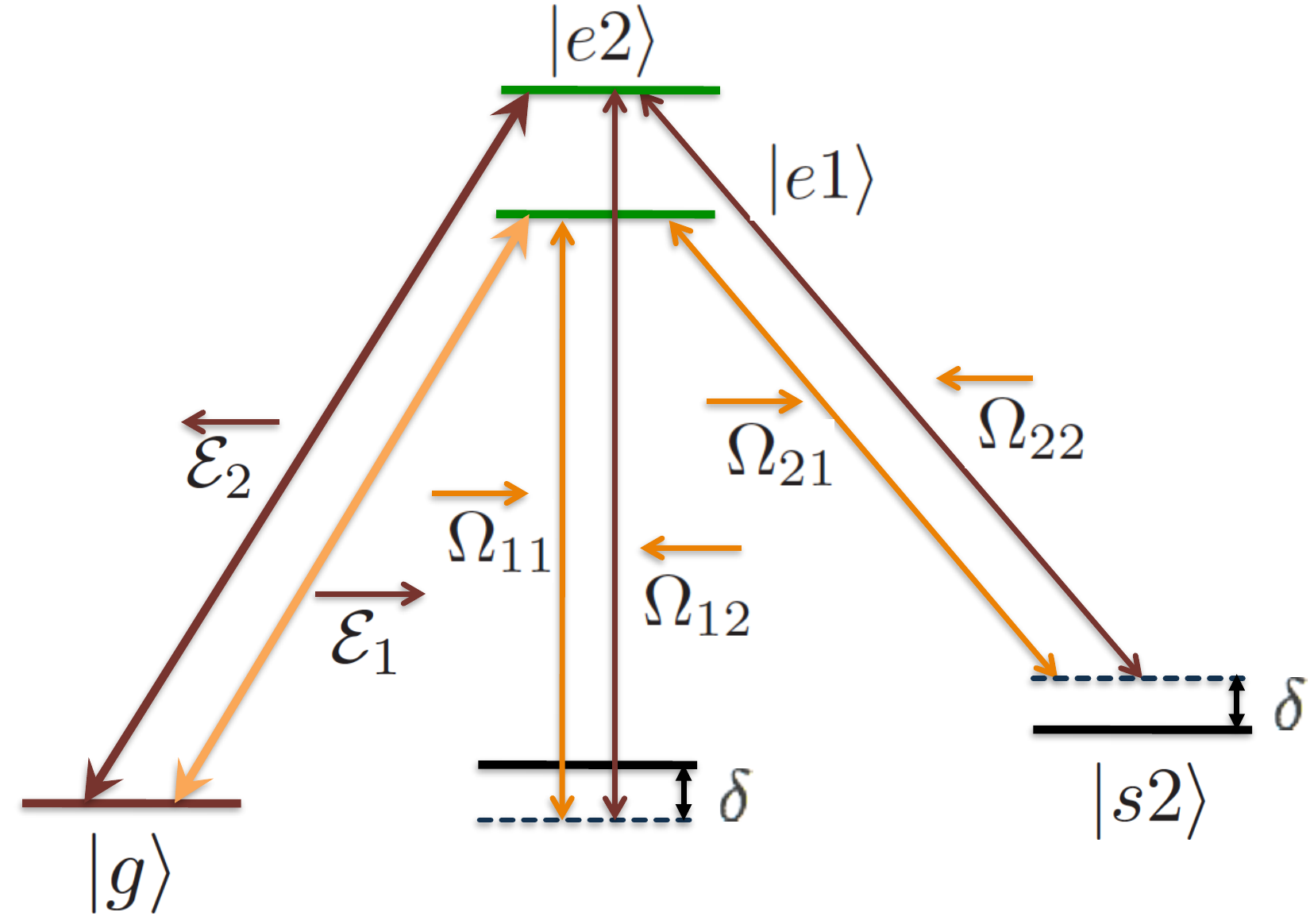}} \\
\end{center}
\caption{ (a) Schematic diagram of the optical waveguide system, interfaced with an ensemble of  atoms where propagating light fields $\mathcal{E}_1$ and $\mathcal{E}_2$ play the role of Dirac spinor components. By adjusting the relevant optical couplings and detunings, the J-R model can be simulated and its topological aspects can be probed by looking at the transmission spectrum. (b) The  level structure  of the interfaced atoms. }
\label{scheme}
\end{figure}

\section{Probing the zero mode in the Jackiw-Rebbi model using photons}

\subsection{Slow-light realization of the J-R model}
The spinor slow light setup we employ to realize the J-R model and subsequently observe the zero-mode is depicted in Fig.~\ref{scheme}. This setup was first proposed by Ruseckas et al. in \cite{Ruseckas}, to which we refer the reader for a detailed explanation of the system and only quote the resulting equation here. The system comprises of a waveguide system coupled to an ensemble of atoms, which could be realized either in a tapered fiber \cite{Nayak,Vetsch10} or a hollow-core waveguide \cite{HCE1,HCE2,HCE3,HCE4,Bajcsy11}. The atoms are characterized by the three hyperfine ground levels; one populated state $|g\rangle$ and two unpopulated states $|s1\rangle$ and $|s2\rangle$. These ground states are coupled to the two excited states $|e1\rangle$ and $|e2\rangle$ by the probe field and control field. The counter-propagating probe beams are described by the electric field amplitudes $E_{1}$ and $E_{2}$, with the respective central frequencies $\omega_{1}$ and $\omega_{2}$ and drive the transitions $|g \rangle \rightarrow |e1 \rangle$ and $|g \rangle \rightarrow |e2 \rangle$. The propagations of the probe beams is controlled by two pairs of counter-propagating control lasers with Rabi frequencies $\Omega_{j1}$ and $\Omega_{j2}$ (with $j = 1, 2$), driving the transitions from the excited states to the unpopulated states. A slowly varying amplitude $\etilde_{j}({\bf r},t)$ is associated with the electric field strength $E_{j}$  of the $j^{th}$ probe field:
\begin{eqnarray}
E_{j}(z,t) = \sqrt{\frac{\hbar \omega}{2 \epsilon}}\etilde_{j}(z,t) e^{- i \omega_{j} t + ik_j z} + c.c. ,
\end{eqnarray}
with $k_1 = \omega_1$ and $k_2 = -\omega_2$, where the speed of light in an empty waveguide is taken to be 1.

The propagation of the slowly varying amplitudes is such that they follow the 1+1 dimensional Dirac-like equation
\begin{align}
\left[ \left( 1+ \frac{1}{v_0}\frac{1}{\sin^2S} \right)\sigma_z - i\frac{1}{v_0}\frac{\cos S}{\sin^2 S} \sigma_y \right]\frac{\partial}{\partial_t}\etilde + \frac{\partial}{\partial_z}\etilde \nonumber \\
 = -\frac{\delta}{v_0\sin S}\sigma_x \etilde.
\end{align}
$v_0 = \Omega^2/g^2 n$ is the group velocity much smaller than 1, where $\Omega/\sqrt{2}$ is the Rabi frequency of the control fields, $g$ is the atom-light coupling strength,  and $n$ is the atomic density. The complex Rabi frequencies are tuned so that $S_{11} = S_{22} = 0$ and $S_{12} = S_{21} = S$ where $\Omega_{ij} = \Omega/\sqrt{2} \exp (iS_{ij})$. $\etilde$ is a column vector of two slowly-varying field components $\etilde_1$ and $\etilde_2$, i.e., $\etilde = (\etilde_1, \etilde_2)^T$ and $\delta$ is the two-photon detuning. In the limit $S=\pi/2$, the above equation reduces to the Dirac equation in 1+1 dimension
\begin{align}
\left( i\partial_t + i v_0\sigma_z\partial_z - \delta\sigma_y \right) \etilde = 0,
\end{align}
which is equal to the Dirac equation introduced in the previous section with the identifications $c = v_0$ and $mc^2/\kappa = \delta$. 

The connection with the J-R model is obvious once we let $\delta$ be spatially varying as $\delta(z) = \delta_{0} \tanh(\lambda z)$. A similar model has been studied in \cite{Unanyan} with a slightly different atomic level scheme and the existence of zero-mode has been briefly commented on, although no connection with the J-R model has been made nor the topological nature of the zero-mode mentioned. By making the connection, it is easily seen that there is  interesting physics to be explored in the slow light system, namely the topologically protected zero mode. Here we discuss how the zero-mode and its topological stability, can be observed experimentally in an feasible set up, based in a driven out-of-equilibrium scenario. 

\subsection{Probing the zero-mode and its topological stability}
\begin{figure}[ht]
\includegraphics[width=3.0cm]{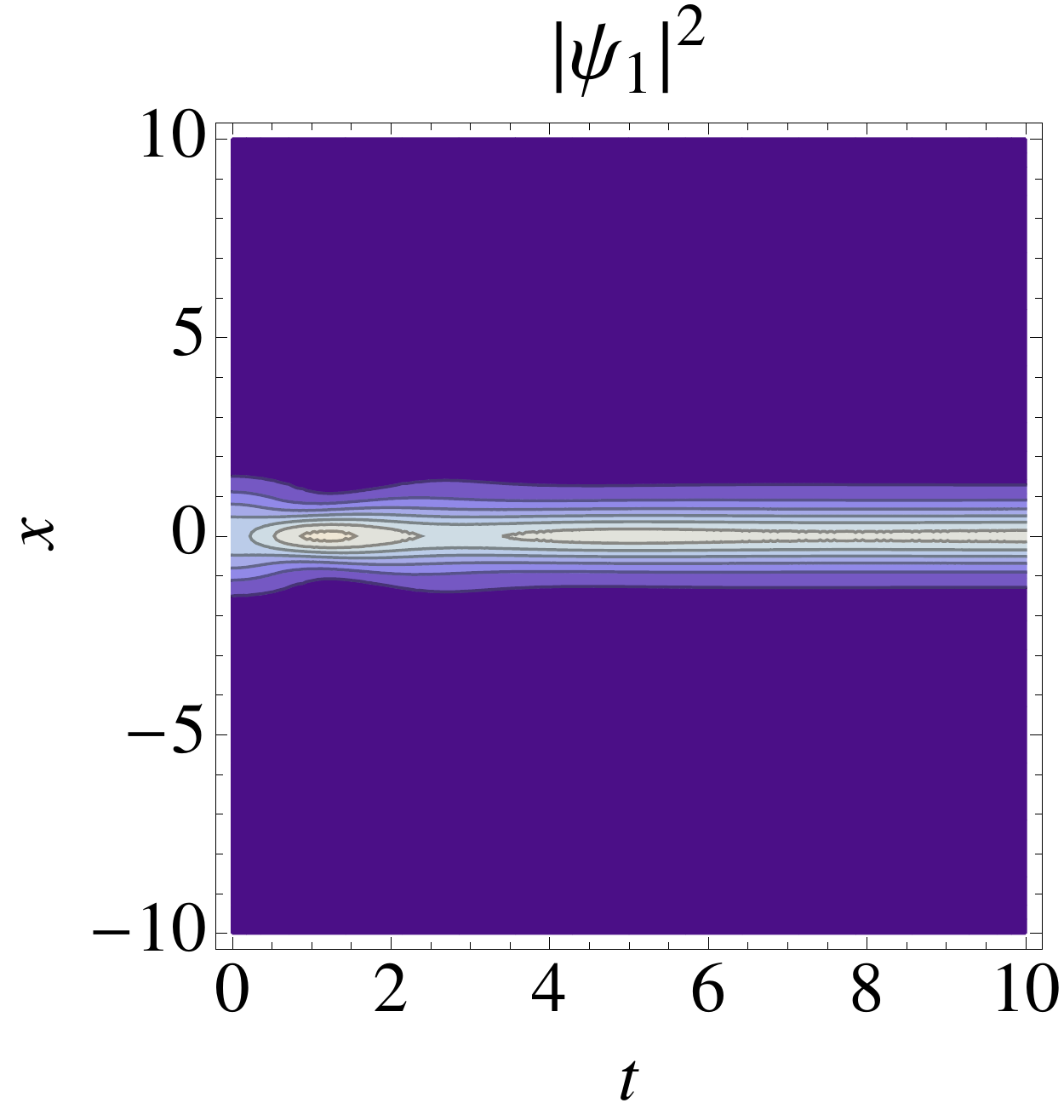}
\includegraphics[width=3.0cm]{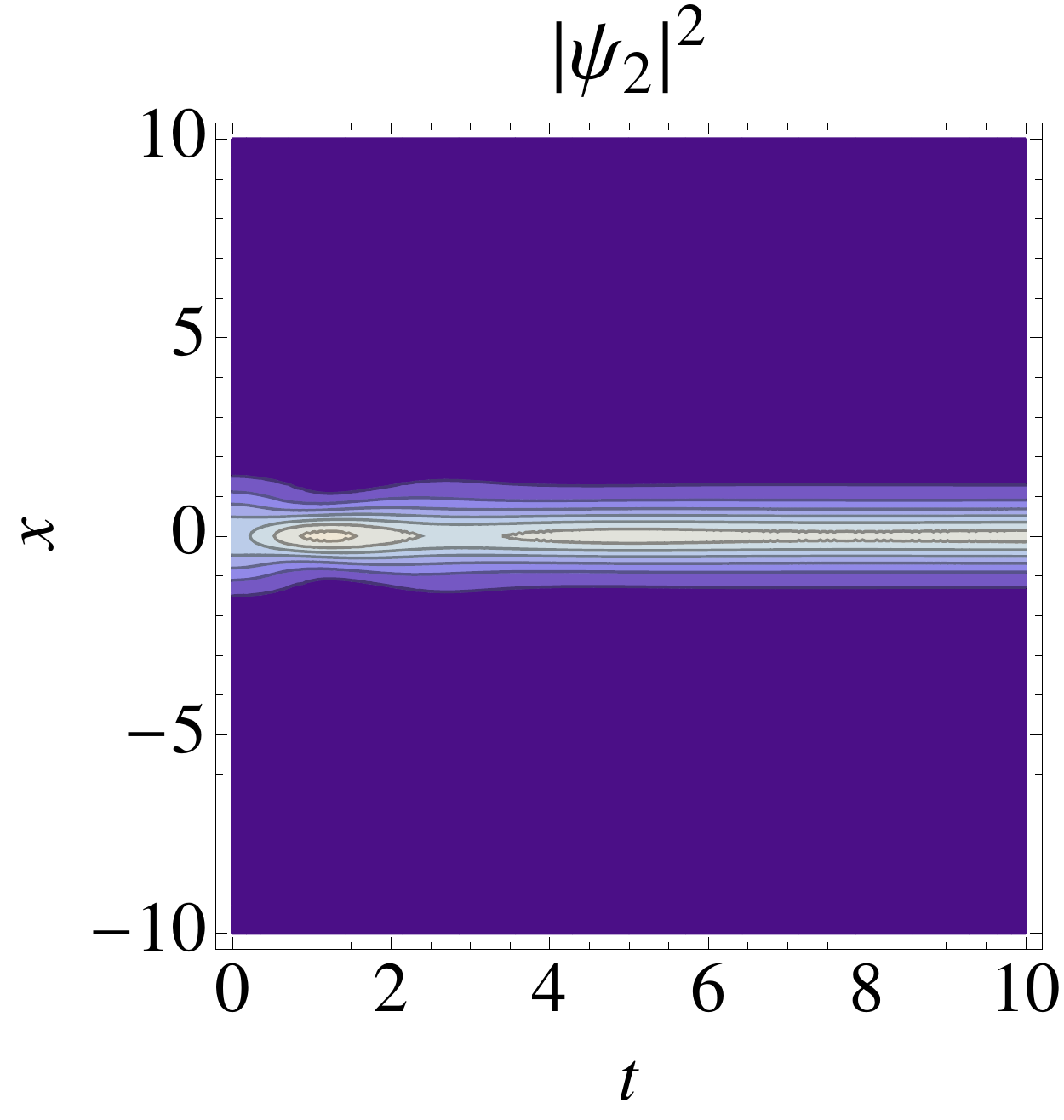}
\includegraphics[width=3.0cm]{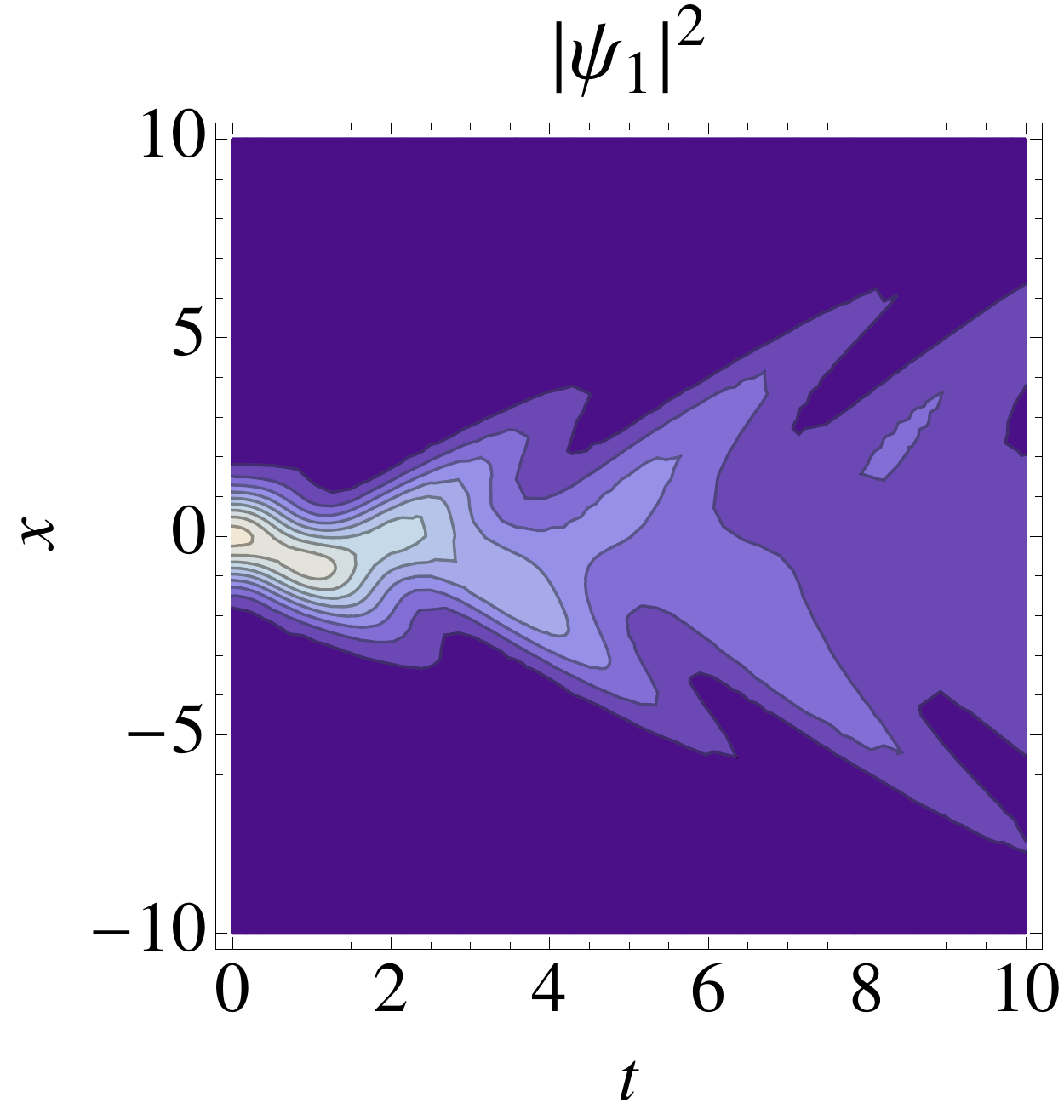}
\includegraphics[width=3.0cm]{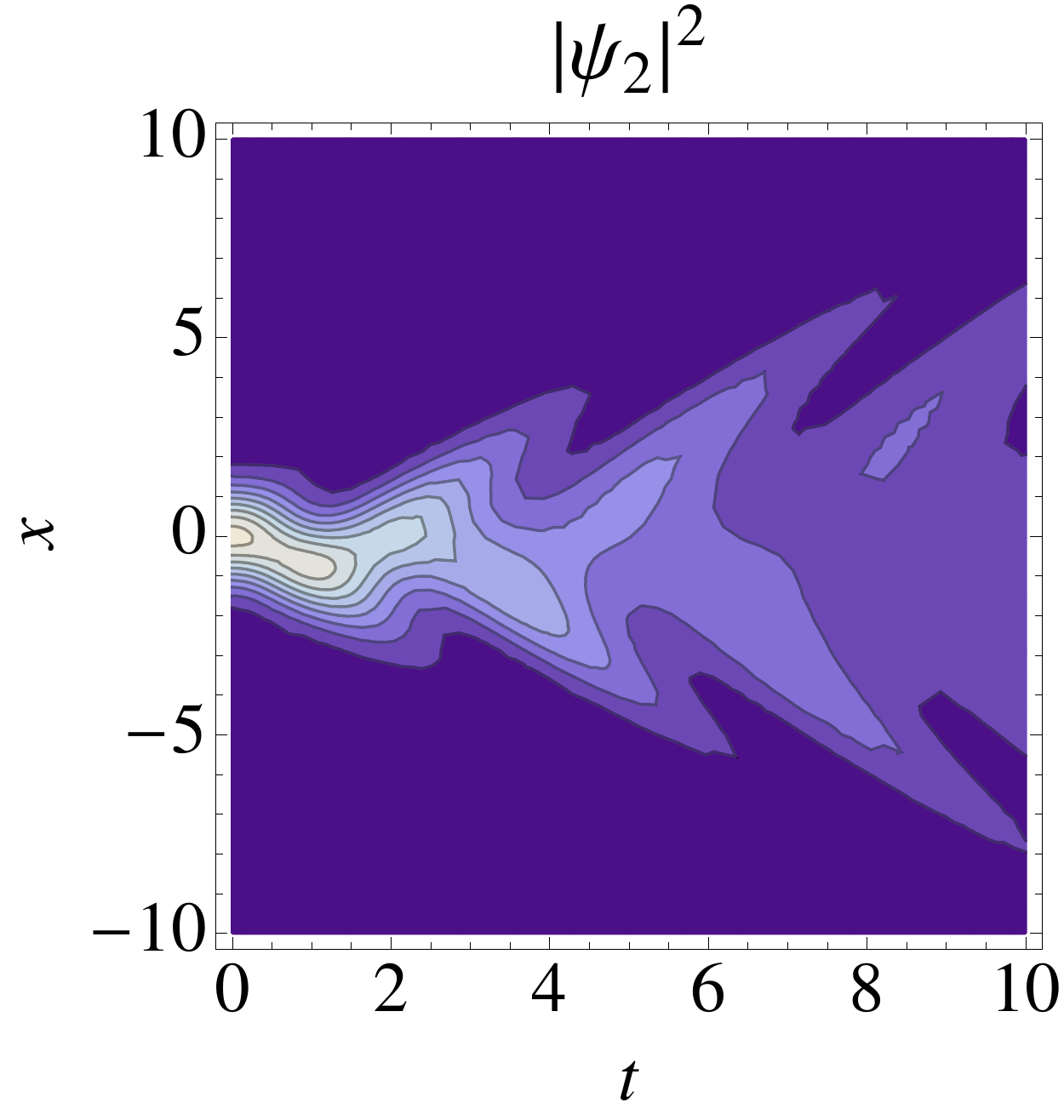}
\caption{Dynamics of an optical gaussian wave packet mimicking the zero-mode wavefunction (\ref{zeromodewf}) in the J-R model (top row), compared with its evolution under the  Dirac equation (bottom row). In the first case the coupling to the background soliton forces the initial wave packet to be trapped, while the second case shows the expected wavefunction spreading. $\psi_1$ and $\psi_2$ corresponds to the top and bottom parts of the spinor, respectively.}
\label{wfevolution}
\end{figure}

Broadly speaking, there are two possible ways to observe the zero-mode in this optical set up.  The first is an adiabatic method, where the light wave packets are adiabatically loaded to prepare an initial state that resembles the zero-mode whose evolution is then observed. In this method, the initial state is prepared by capturing an initial pulse in the medium via the usual electromagnetically induced transparency (EIT) way, i.e., by slowly turning off the forward traveling control fields which are initially on\cite{Lukinreview}. Then, all the control fields are slowly turned on again, including the coupling fields $\Omega_{12}$ and $\Omega_{21}$. At this point, the pulses are trapped as stationary light and go through the effective evolution governed by the Dirac equation. The dynamics of the spinor fields then differ significantly, depending on the initial condition of the wave packets as shown in Fig.~\ref{wfevolution}, where we have solved the Dirac equation with the initial gaussian wave packet Exp$[-x^2/2\sigma^2]/\sqrt{\sqrt{\pi}\sigma]}$ with $\sigma=1.2$ to mimic the zero-mode spinor. 

\begin{figure}[ht]
\begin{center}
\subfigure[$ $]{
\includegraphics[scale=0.15]{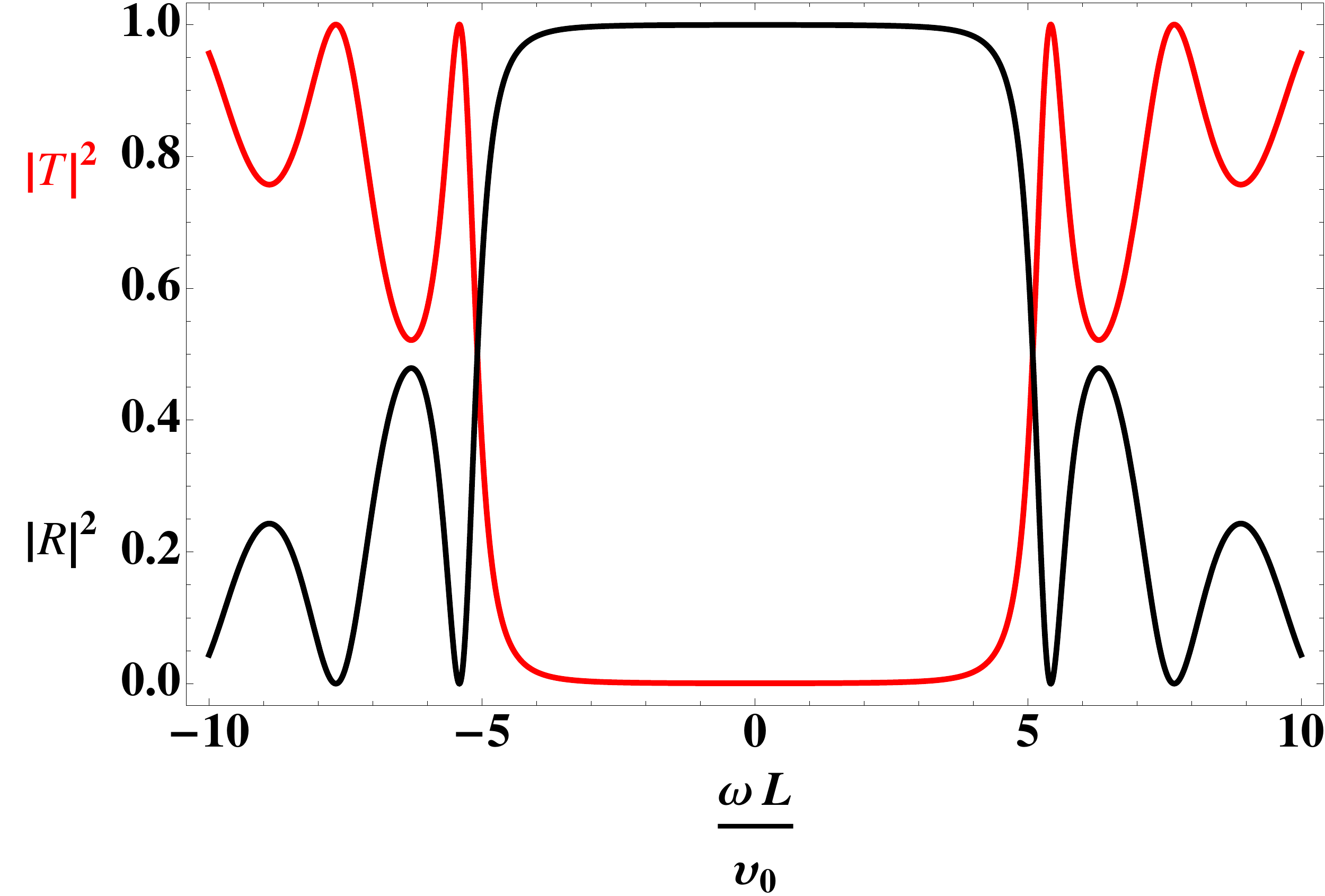}}
\subfigure[$ $]{
\includegraphics[scale=0.15]{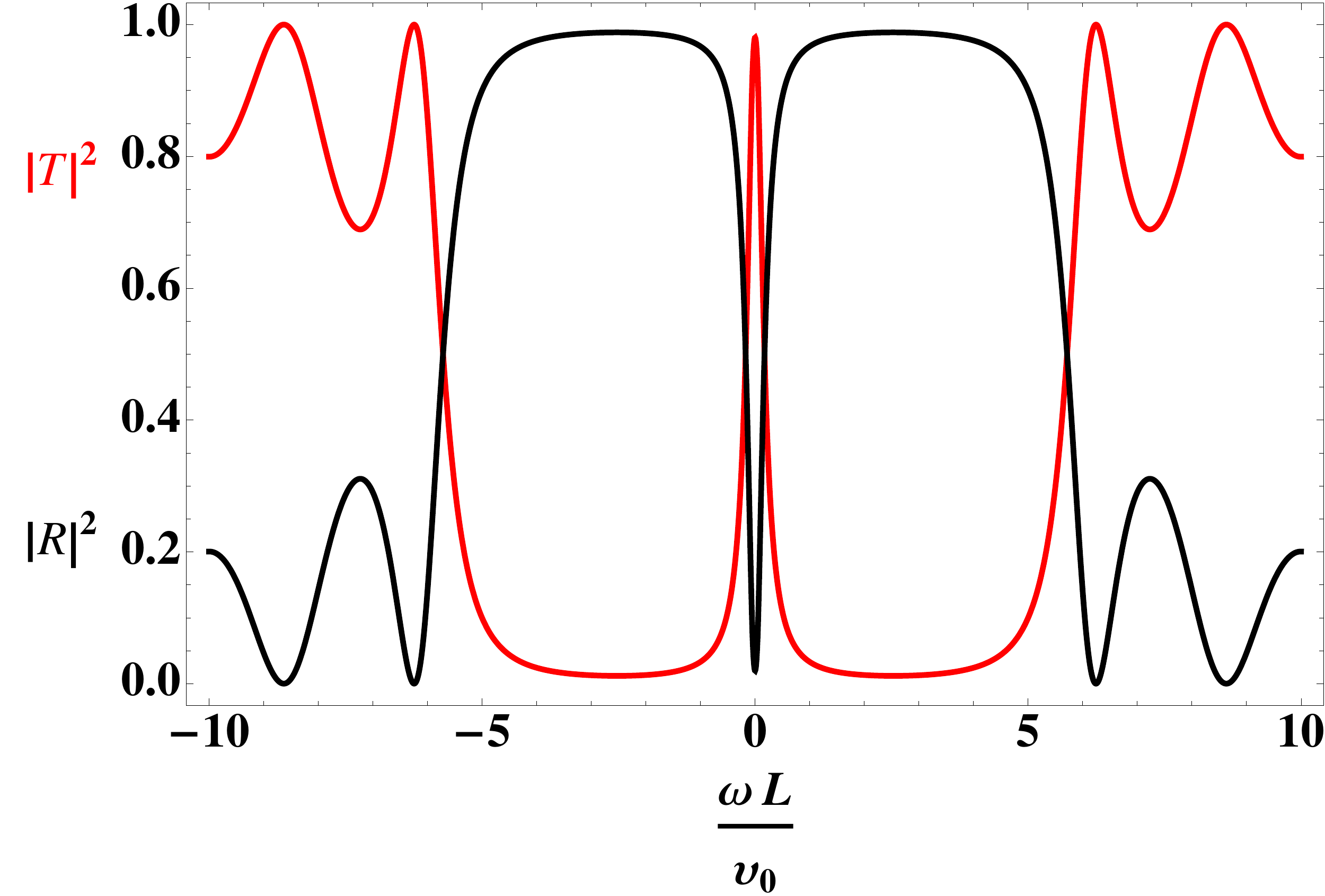}}
\caption{The reflection $|R|^{2}$ (black) and transmission $|T|^{2}$ (red) curves for the effective Dirac particle (a) without the soliton background and (b) with a soliton field whose profile is 0.25 tanh$(0.02 z)$. (a) shows the Dirac mass bandgap whereas (b) shows near-unity transmission near the zero-energy due to the bound zero-mode.}
\label{zero-mode}
\end{center}
\end{figure}

The second method, better suited for this particular experimental realization and also easier to implement, is to look at the transmission and reflection of an incident probe field $\etilde_1(z=0)$. Similar studies have been carried out quite recently  using  transmission to probe strongly interacting effects in similar polaritonic systems \cite{Hafezi, Petrosyan,Das}. Consider a monochromatic probe field $\etilde_1 = \alpha \,e^{-i \Delta \omega t}$ impinging from the left, while $\etilde_2 = 0$. One can study the transmission and reflection spectrum of this field where the transmitted field will come out to the right of the waveguide as $\etilde_1(z=d) = \alpha T \, e^{-i \Delta \omega t}$, whereas the reflected part is $\etilde_2(z=0) = \alpha R \,e^{- i \Delta \omega t}$ as shown in Fig.~\ref{scheme}(a). The constant mass case can be solved analytically \cite{Ruseckas} and shows the behaviour depicted in Fig.~\ref{zero-mode}(a). There is a finite window of perfect reflection due to the well known band gap proportional to the mass energy. Upon introducing the spatially dependent mass term discussed earlier, i.e. $\tanh (\lambda z)$, transmission reappears at the center of the bandgap as shown in Fig.~\ref{zero-mode}(b), due to the existence of the zero-mode. Here and below, we follow \cite{Ruseckas} and assume that the size of atomic length cloud interacting with the propagating light is $L = 300 \mu m$ and the latter's group velocity $v_{0} = 17 m/s$. For these values, the maximum two photon detuning $\delta_0 = 0.25v_0/L$ lies well within the EIT transparency window.

\begin{figure}[ht]
\begin{center}
\vspace{0.5cm}
\subfigure[$ $]{
\includegraphics[scale=0.15]{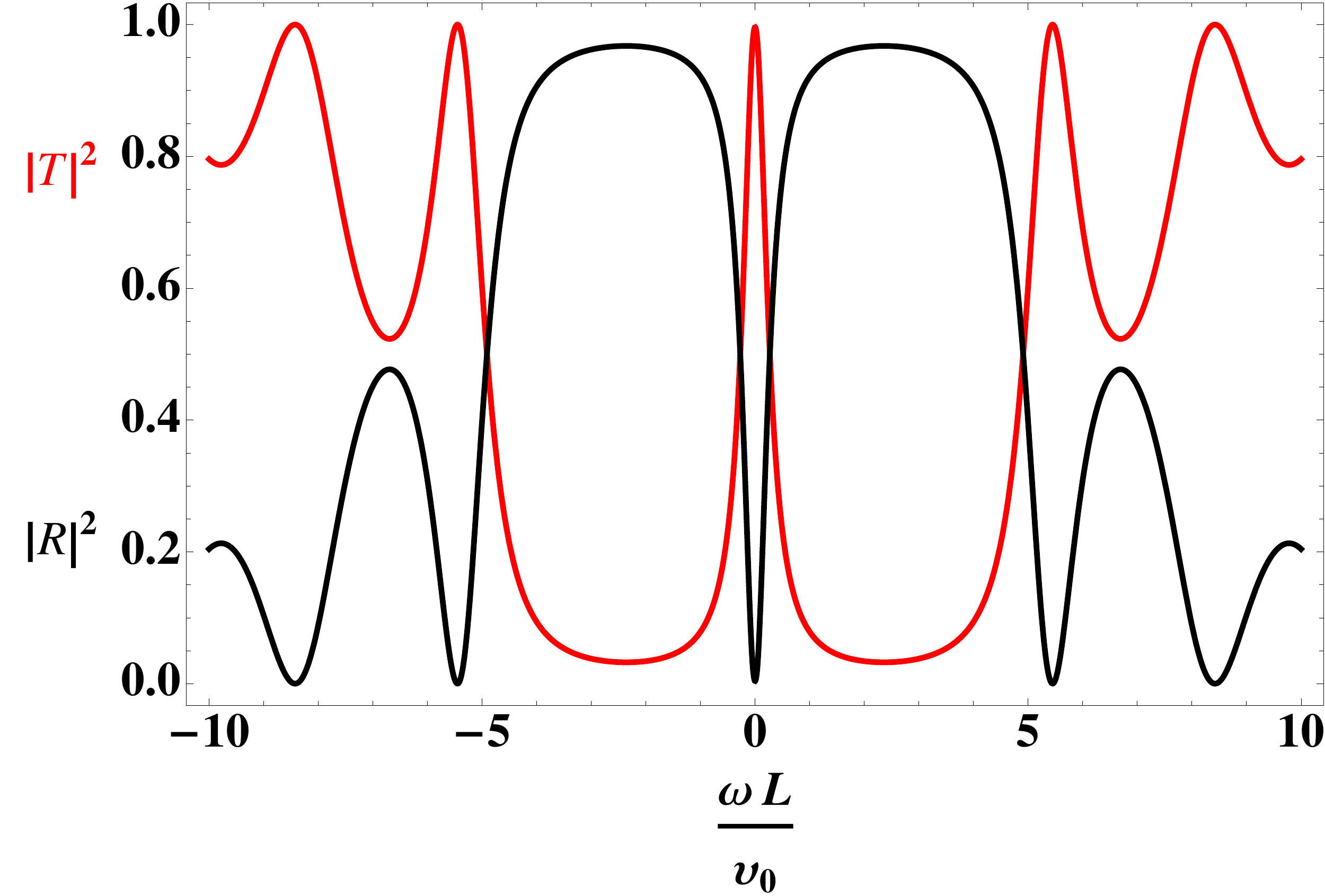}}
\subfigure[$ $]{
\includegraphics[scale=0.15]{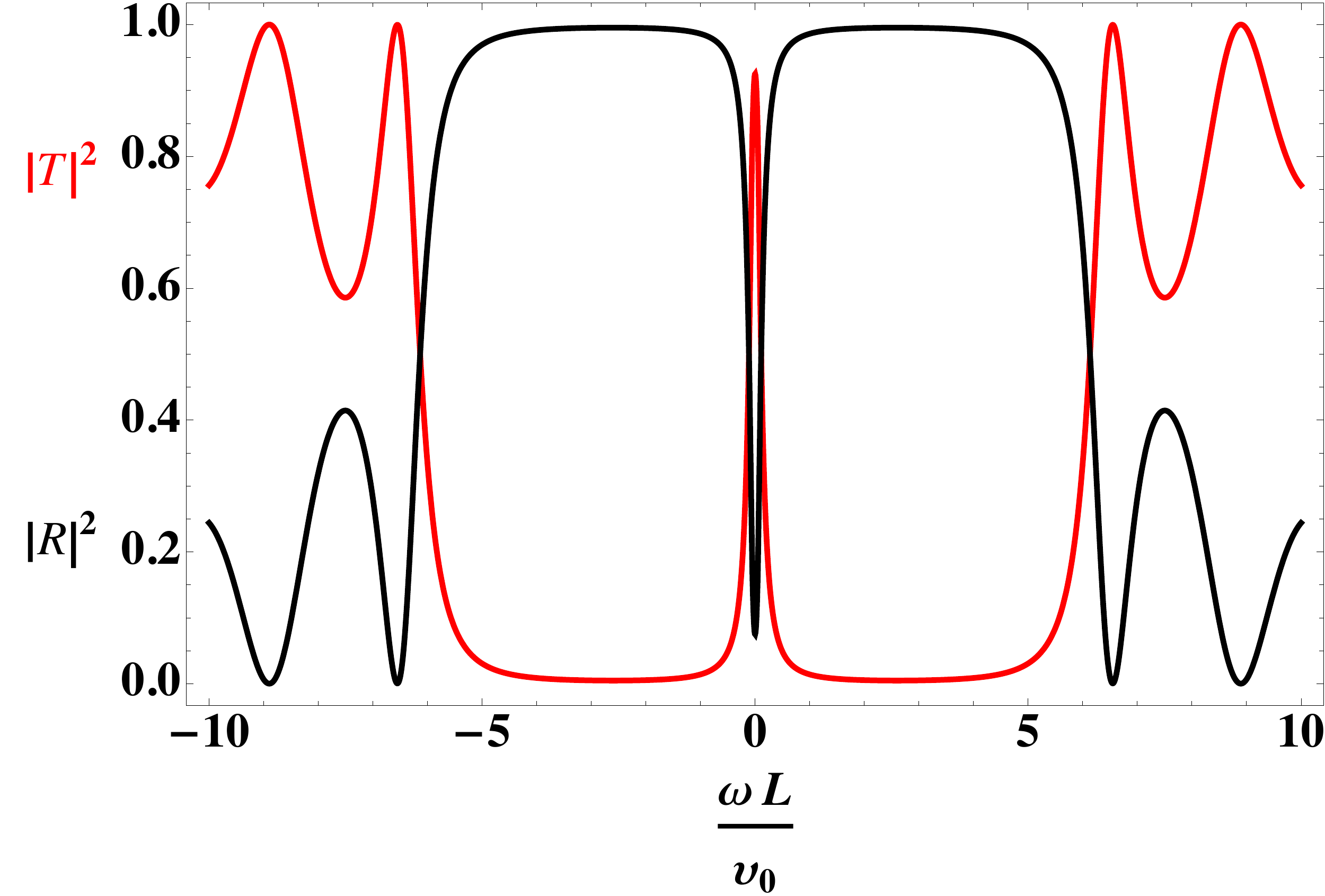}}
\subfigure[$ $]{
\includegraphics[scale=0.15]{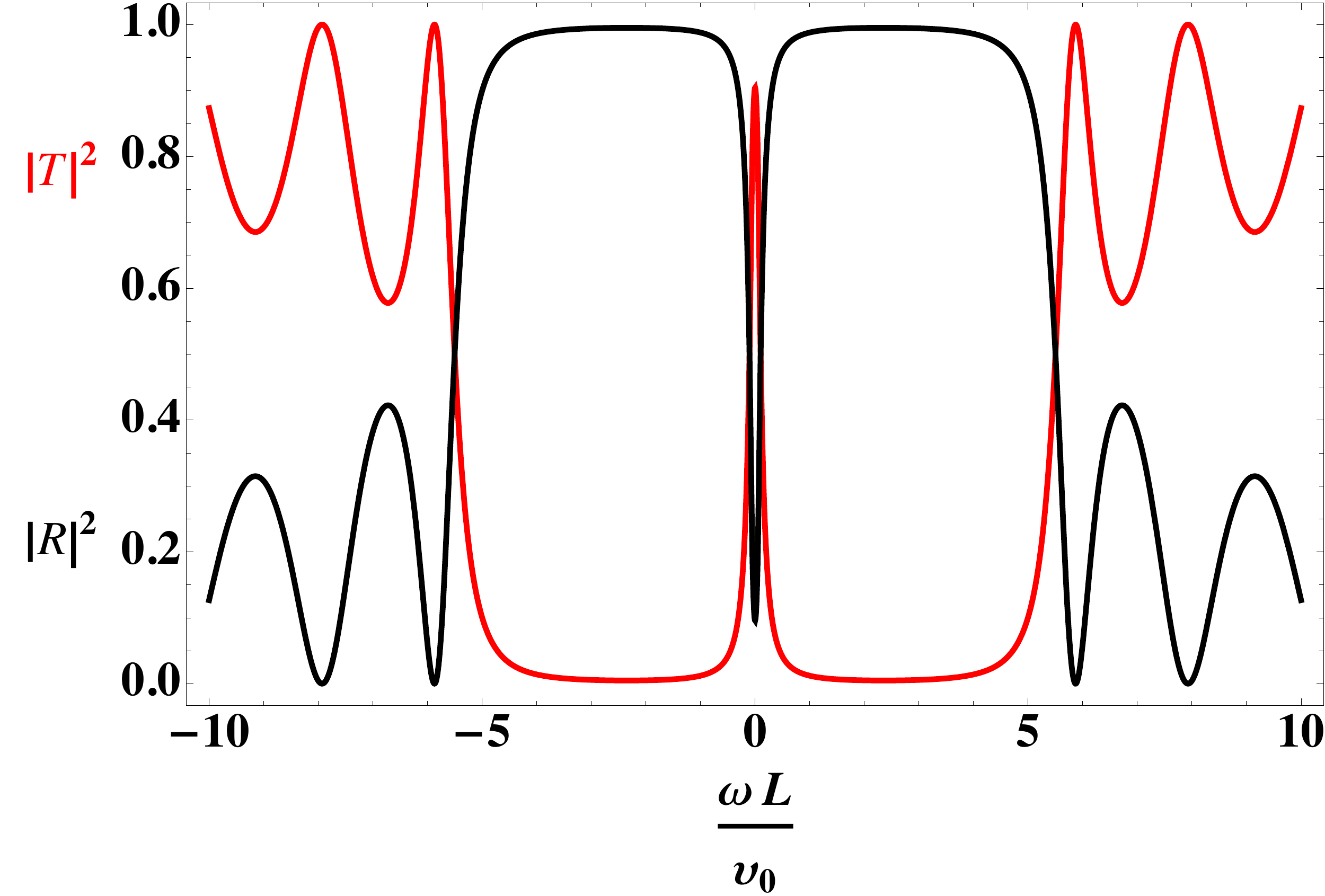}}
\caption{The transmission spectrum showing the stability of the zero-mode under various types of perturbations. Topological protection under (a) a change in the soliton profile to a sine function $0.25 \sin (0.01 z)$, (b) random fluctuations up to $30\%$ in the soliton field used in Fig.~\ref{zero-mode}. The robustness against experimental errors in engineering the model Hamiltonian is shown in (c) where there is a correction to the Dirac equation through a the mixing angle: $\Delta S=0.2\pi/2$. }
\label{effects of noise}
\end{center}
\end{figure}

Using this second method, the topological nature of the zero-mode can be readily probed by looking at the transmission spectrum while perturbing the mass profile. The latter can be done by tuning the two photon detunings using standard optical methods, like AC stark shifting used in slow light experiments\cite{ Lukinreview,Marangosreview}. For example, changing the hyperbolic tangent function to the sine function while preserving the topology of the profile, i.e., the mass term takes the value of -$\delta_0$ and +$\delta_0$ at $z=0$ and $z=L$, respectively, has little effect on the transmission at the center of the bandgap as shown in Fig.~\ref{effects of noise}(a). Changing the hyperbolic tangent function to a sine function that goes through 0 in the middle of the waveguide and has $\pm \delta_0$ at $z=0,L$ has qualitatively the same transmission spectrum. The zero-mode is also protected from random fluctuations in the mass profile as shown in Fig.~\ref{effects of noise}(b), where we have assumed random fluctuations within 30\% of $\delta(z)$, i.e., $\delta(z) = \delta_0 \tanh (0.02z)(1+\epsilon (z))$ where $\epsilon(z)$ is a uniform random number in the interval $[-0.3,0.3]$ with the resolution of 0.1$\mu m$. As a final type of perturbation we note that experimentally, it might be difficult to set the phases of the control fields exactly to the required value of the mixing angle $S = \pi/2$ in Eq.~$(6)$. To study the effect of this imperfection we have set $S = \pi/2 + \Delta S$ where $\Delta S$ takes 20\% of the desired value, the result of which is depicted in Fig.~\ref{effects of noise}(c). The presence of the zero-mode persists upon experimental errors in creating the exact Hamiltonian (see supplementary information for further details).

%
%

\section{Conclusions}

In this work, we have shown that it is possible to simulate the Jackiw-Rebbi  model and probe its topological nature in a driven slow-light based system using current technology. By introducing spatially-dependent optical detunings, it is possible to simulate the Dirac equation with a spatially dependent static soliton field, and allows one to directly probe the topologically protected zero mode. The robustness of this zero-mode can be tested by changing the spatial profile of the detunings, while continuously observing the optical transmission spectrum of the system. 

Before closing we would like to also briefly comment on other types of models that might be studied in the same system in future works. Firstly, adding interactions in the system through EIT photon nonlinearities could allow the study of interesting strongly correlated physics. Intra-species repulsion would make the bosons behave like fermions in many ways, and it would be interesting to think about how this affects the (lack of) charge fractionalization effect  in bosons. Interacting random mass Dirac model is another interesting possibility as studying the interplay between interactions and randomness is an important field being actively studied.  Yet another possibility is to think of the mass term as the Lorentz-scalar potential which can act as a confining potential, given a proper spatial dependence \cite{Critchfield,Fishbane}. This model has been shown to act as a phenomenological model of quark confinement motivated by the `MIT-bag' model \cite{Chodos}.

We would like to acknowledge the financial support provided by the National Research Foundation and Ministry of Education Singapore (partly through the Tier 3 Grant ``Random numbers
from quantum processes"), and travel support by the EU IP-SIQS.

\section*{Supplementary information}

Here we present for reference  further results in relation to the robustness of the set up and the stability of the zero mode against possible experimental implementation errors. This is done for a wider parameter range than in the main text. In the first figure, as in the main text Fig. $ 5(c)$, we discuss errors and its effect in implementing the required relative phase between the optical fields. In the second figure we discuss the effect of fluctuations in the optical detunings profiles (as in main text Fig. $5 (b))$.
\begin{figure}[ht]
\begin{center}
\subfigure[$ $]{
\includegraphics[scale=0.15]{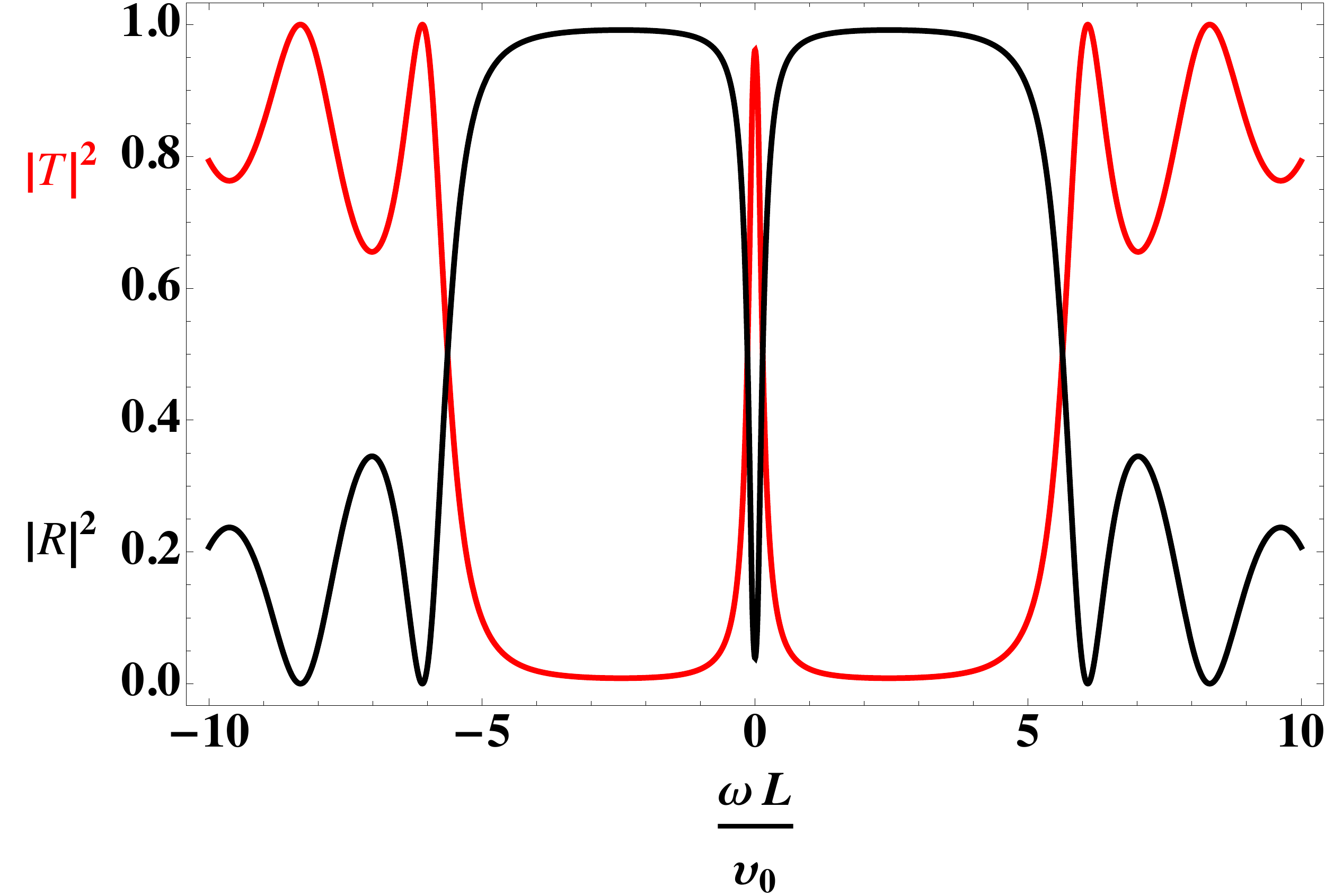}}
\subfigure[$ $]{
\includegraphics[scale=0.15]{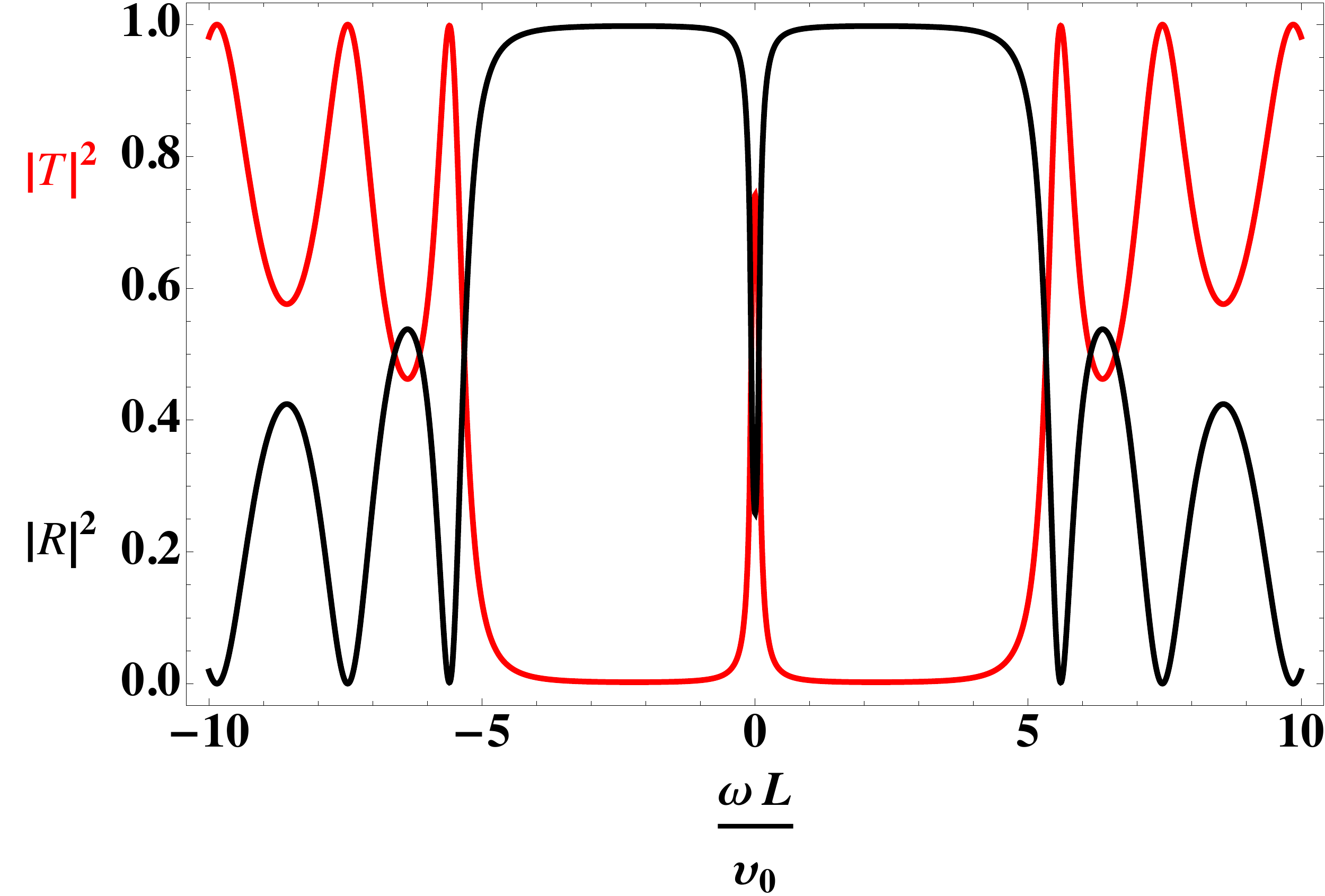}}
\subfigure[$ $]{
\includegraphics[scale=0.15]{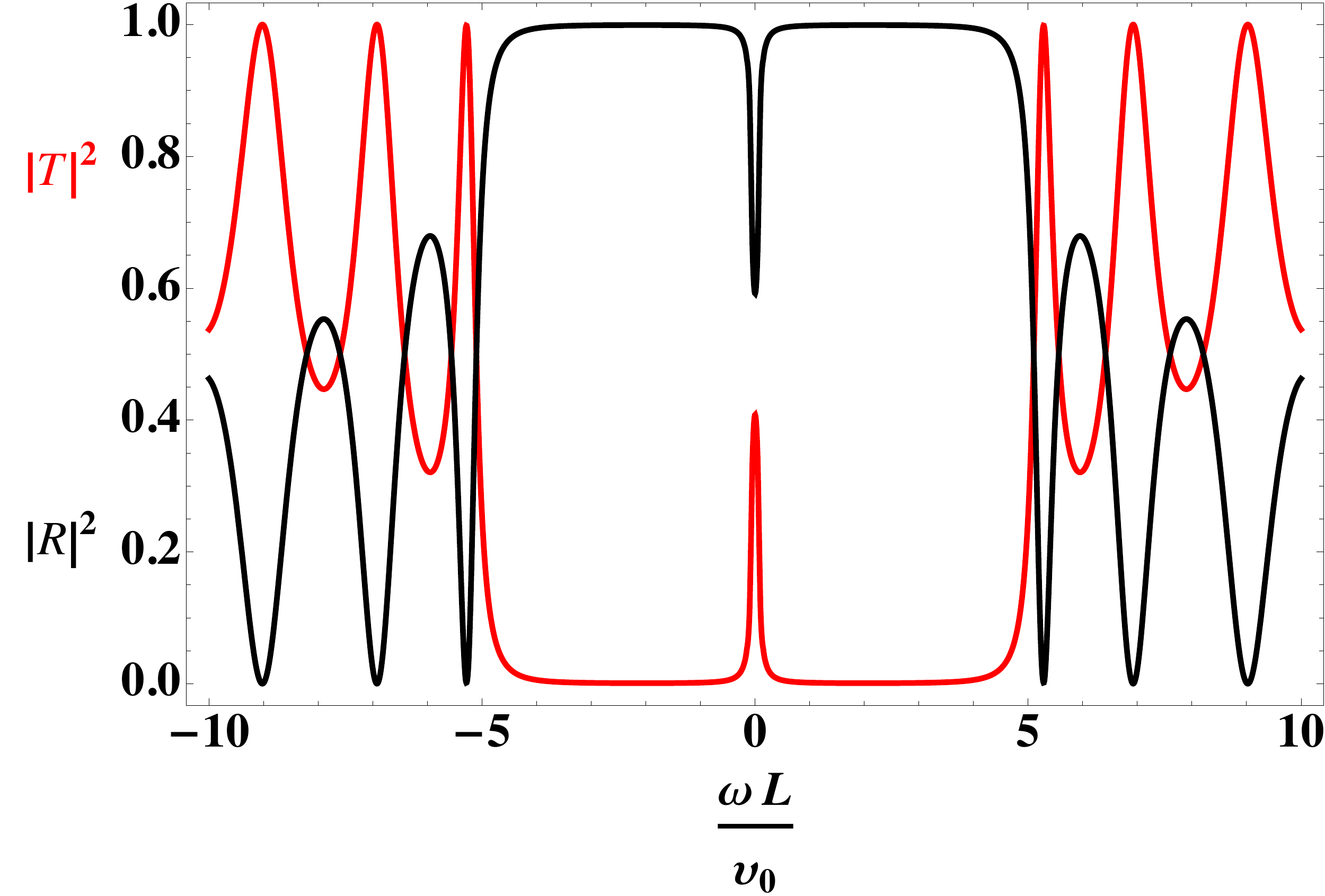}}
\subfigure[$ $]{
\includegraphics[scale=0.15]{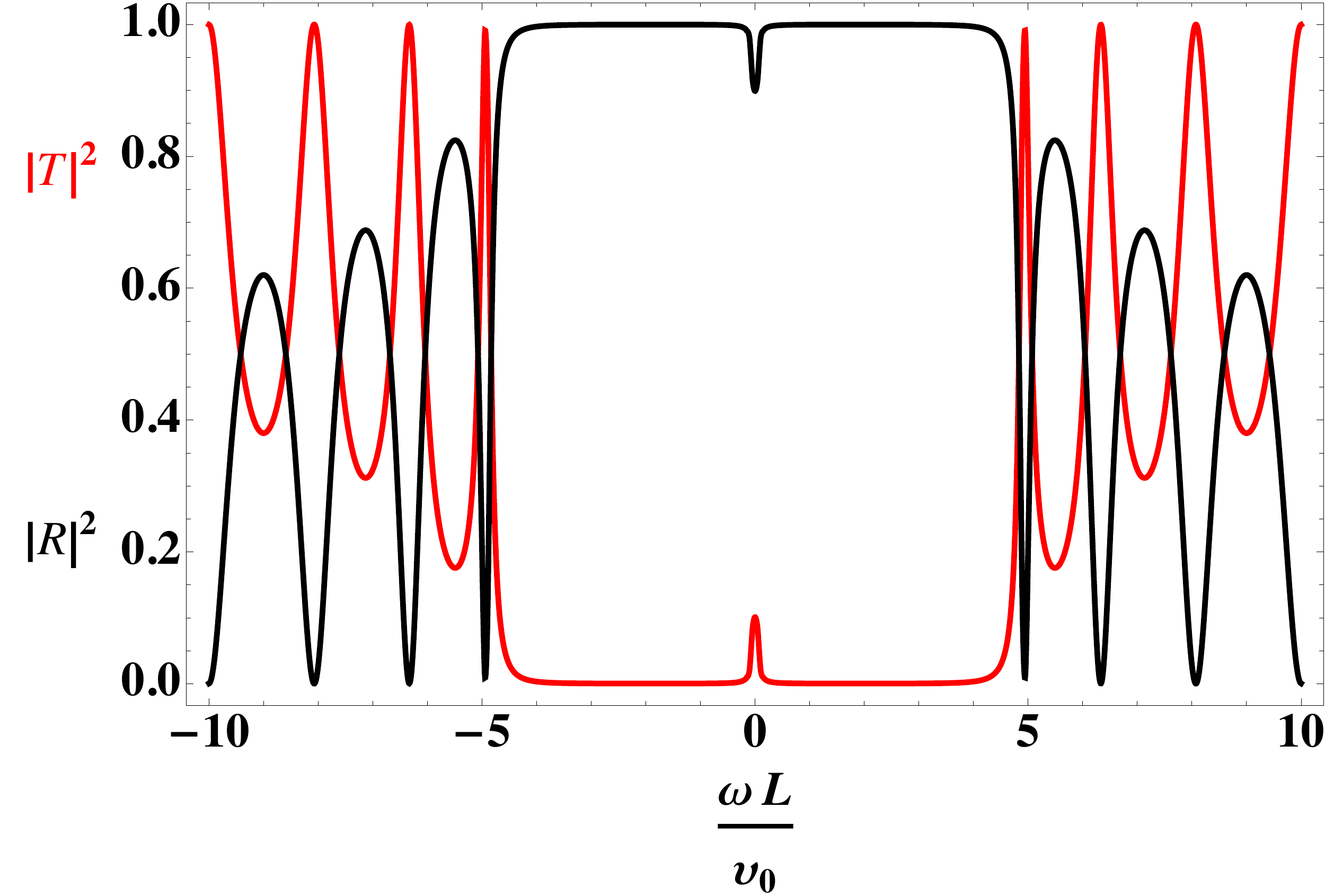}}
\caption{Departure from the Dirac equation for different values of the mixing angle (see Eq. (6) ) for $\Delta S$ = 0.1,  0.3, 0.4, and 0.5 from (a) to (d).  Dirac dynamics persists up to values of  0.3-(subfigure b). }
\end{center}
\end{figure}

\begin{figure}[ht]
\begin{center}
\subfigure[$ $]{
\includegraphics[scale=0.15]{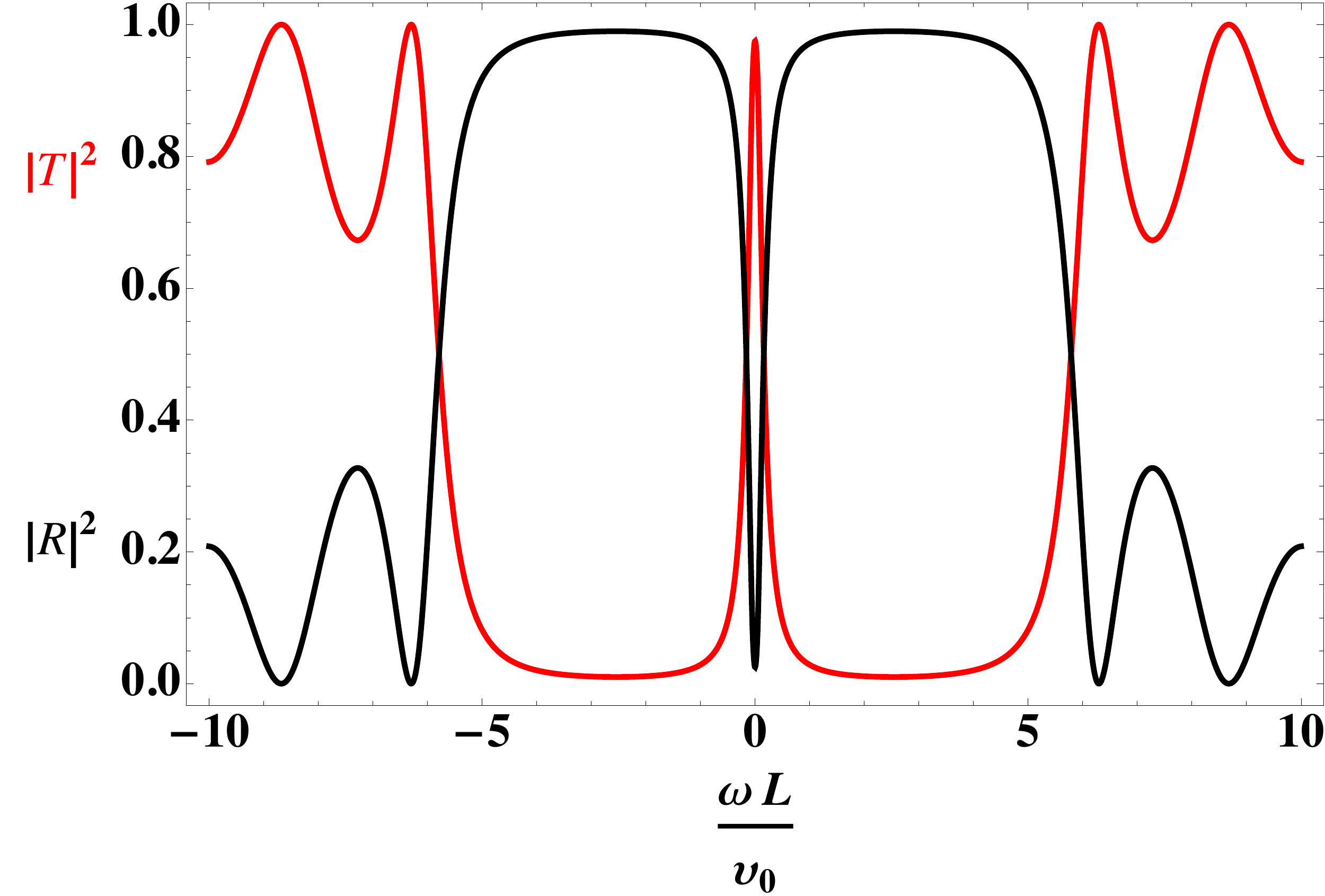}}
\subfigure[$ $]{
\includegraphics[scale=0.15]{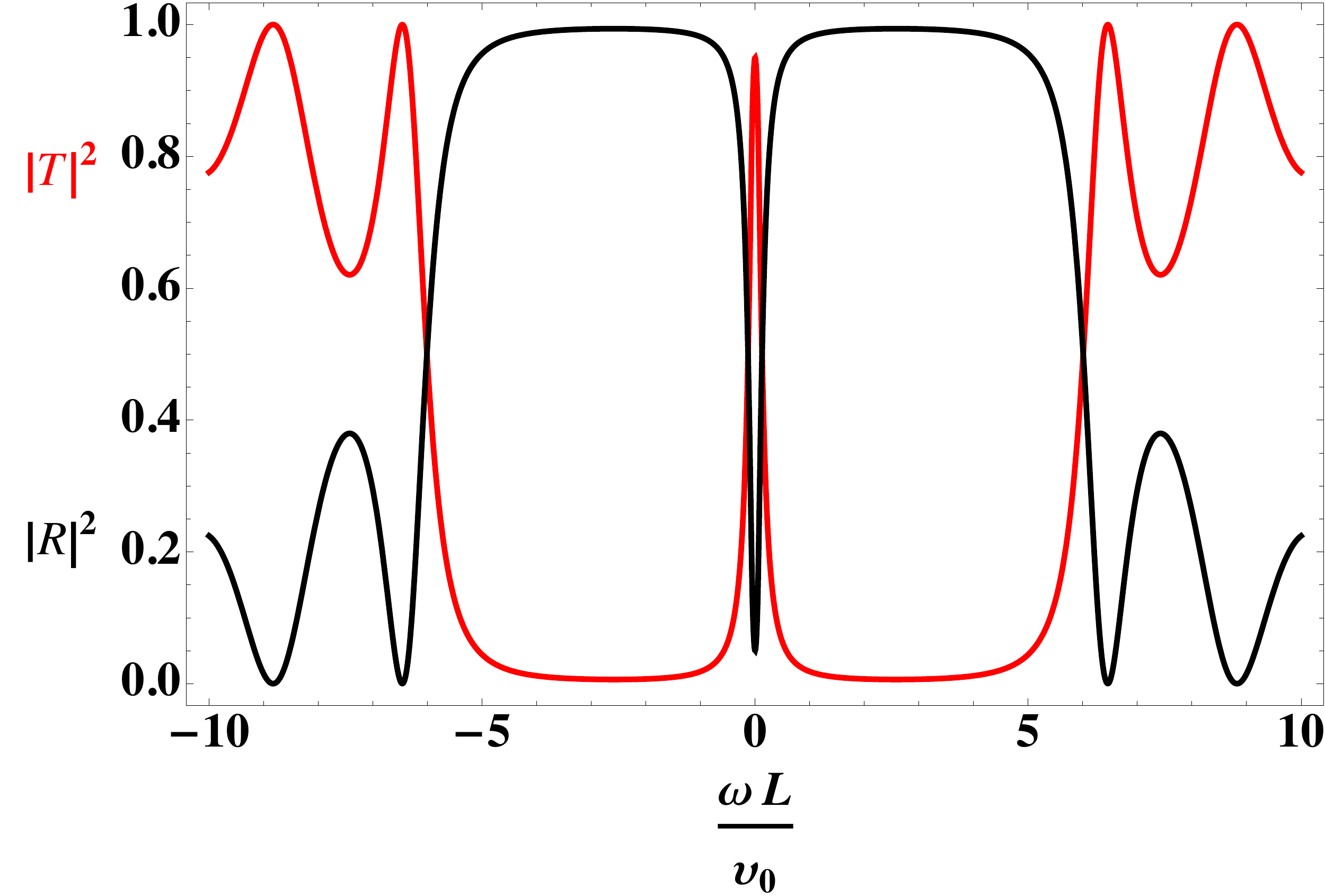}}
\subfigure[$ $]{
\includegraphics[scale=0.15]{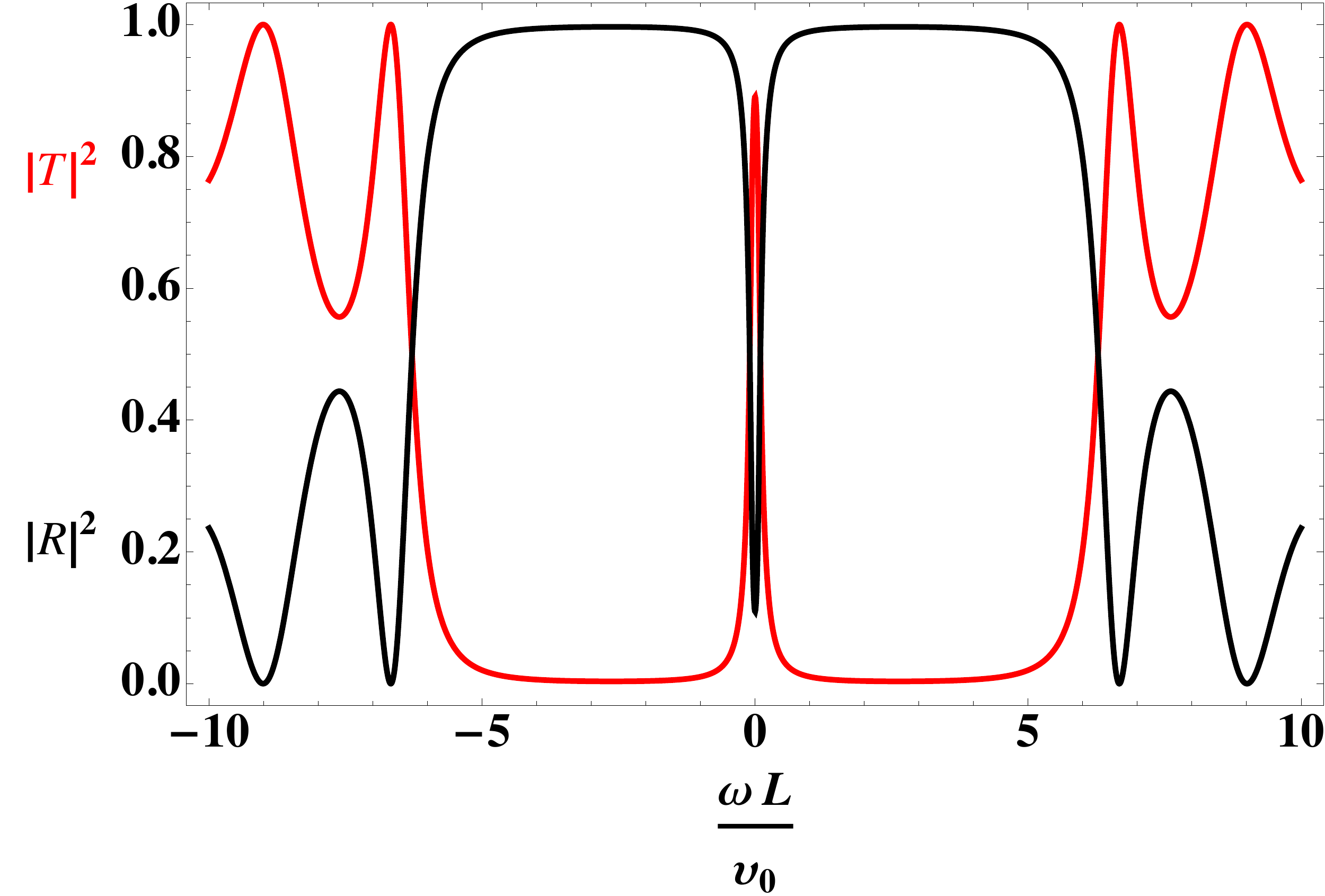}}
\subfigure[$ $]{
\includegraphics[scale=0.15]{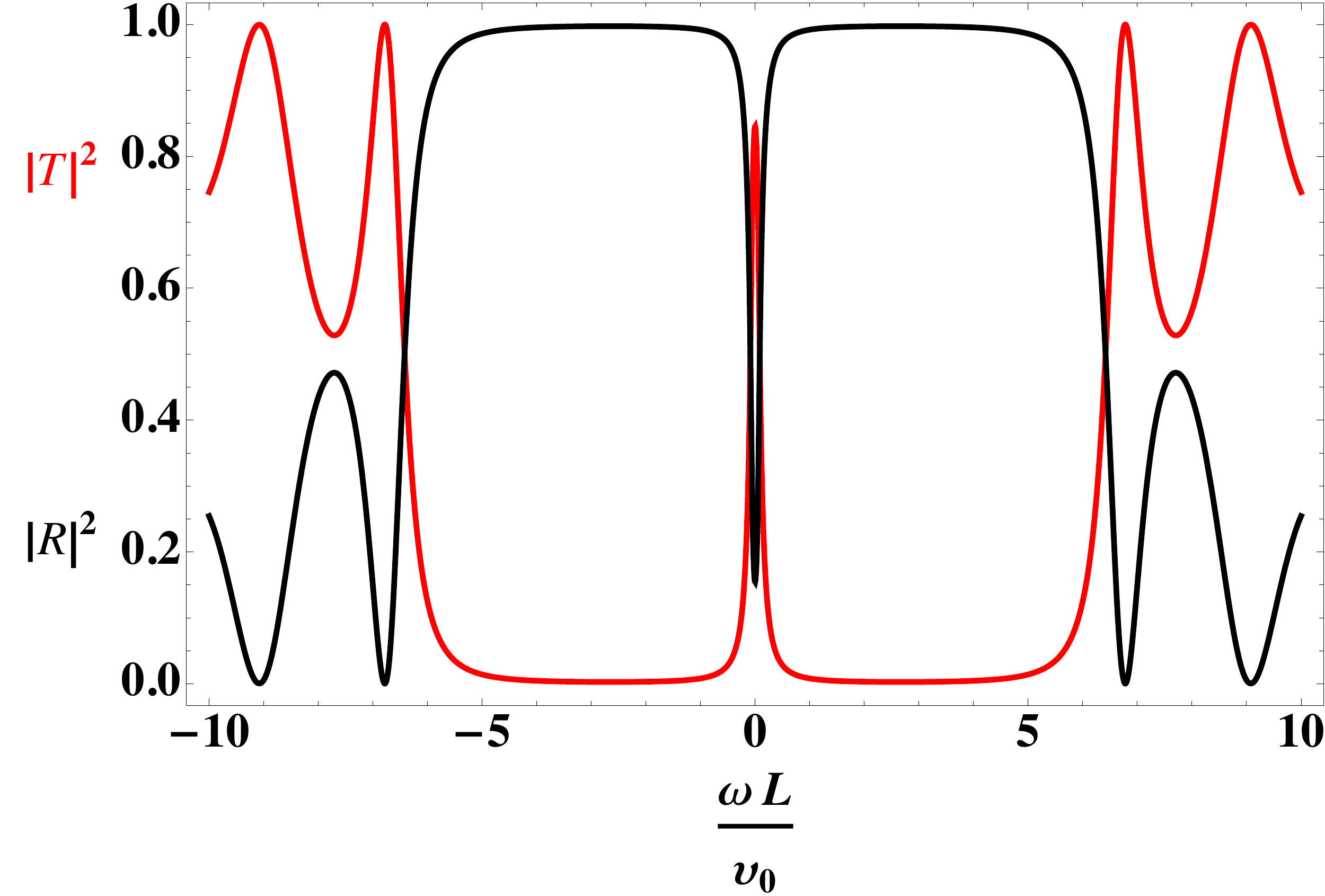}}
\caption{Stability of the zero mode under random fluctuations in the soliton (implemented via the two photon detunings) profile. $\epsilon (z)$ = 5, 20, 40, and 50 $\%$ from (a) to (d). We see that the zero mode remains there in-spite relative large values of fluctuation although the maximum possible transmission is gradually reduced from roughly $95\%$ to $85\%$.}
\end{center}
\end{figure}


\begin{thebibliography}{0}
 \bibitem{CiracZoller2012a} I. Cirac and P. Zoller, Nat. Phys.~\textbf{8}, 264 (2012).

\bibitem{RoosBlatt2012} R. Blatt and C. Roos, Nat. Phys.~\textbf{8} 277 (2012).

\bibitem{BlochNascimbene2012} I. Bloch, J. Dalibard, S. and Nascimbene, Nat. Phys.~\textbf{8} 267 (2012).

\bibitem{reviews} A. Tomadin and R. Fazio, J. Opt. Soc. Am. B. \textbf{27} A130 (2010); M. Hartmann, F. G. S. L. Brand\~ao and M. B. Plenio, Laser Photonics Rev.~\textbf{2} 849 (2008);A. Houck, H. Tureci, and J. Koch, Nat. Phys.~\textbf{8} 292 (2012).

\bibitem{SIP2} D. G. Angelakis, M. F.  Santos, and S. Bose, Phys. Rev. A \textbf{76} 031805(R) (2007).

\bibitem{SIP1} M. J. Hartmann, F. G. S. L. Brand\~ao, and M. B. Plenio, Nat. Phys.~\textbf{2} 849 (2006).

\bibitem{SIP3} A. D. Greentree, C. Tahan, J. H. Cole, and L. C. L. Hollenberg, Nat. Phys.~\textbf{2} 856 (2006).

\bibitem{RossiniFazio07}
D. Rossini and R. Fazio Phys. Rev. Lett.~\textbf{99}, 186401 (2007).

\bibitem{ChoBose08} J. Cho, D.G. Angelakis,and  S. Bose, Phys. Rev. Lett.~\textbf{101}, 246809 (2008).



\bibitem{Lukinreview} M. D. Lukin, Rev. Mod. Phys.~\textbf{75}, 457 (2003).

\bibitem{Marangosreview} M. Fleischhauer, A. Imamoglu, and J. P. Marangos, Rev. Mod. Phys.~\textbf{77}, 633 (2005).

\bibitem{Bajcsy2009} M. Bajcsy, S. Hofferberth, V. Balic, T. Peyronel, M. Hafezi, A. S. Zibrov, V. Vuletic, and M. D. Lukin, Phys. Rev. Lett.~\textbf{102}, 203902 (2009).

\bibitem{Chang} D. E. Chang, V. Gritsev, G. Morigi, V. Vuleti\'c, M. D. Lukin, and E. A. Demler, Nat. Phys.~\textbf{4}, 884 (2008).

\bibitem{Angelakis11} D. G. Angelakis, M.-X. Huo, E. Kyoseva, and L. C. Kwek, Phys. Rev. Lett.~\textbf{106}, 153601 (2011).

\bibitem{Angelakis} D.G. Angelakis, M. Huo, D. Chang, L.C. Kwek, and V. Korepin, Phys. Rev. Lett.~\textbf{110}, 100502 (2013).

\bibitem{carusotto2009fermionized}
I.~Carusotto, D.~Gerace, H.~Tureci, S.~De~Liberato, C.~Ciuti, and
  A.~Imamo{\u{g}}lu, Phys. Rev. Lett. \textbf{103}, 33601
  (2009).
  
\bibitem{grujic2012angelakis}
T.~Grujic, S.~R. Clark, D.~Jaksch, and D.~G. Angelakis, New Journ. of Phys., \textbf{14}, 103025
  (2012).

\bibitem{Lamata}  L. Lamata, J. Le\'on, T. Sch\"atz, and E. Solano, Phys. Rev. Lett.~\textbf{98}, 253005 (2007).

\bibitem{Gerritsma}  R. Gerritsma, G. Kirchmair, F. Z\"ahringer, E. Solano, R. Blatt, and C. F. Roos, Nature \textbf{463}, 68 (2010).

\bibitem{Casanova}  J. Casanova, J. J. Garc\'ia-Ripoll, R. Gerritsma, C. F. Roos, and E. Solano, Phys.~Rev.~A \textbf{82}, 020101(R) (2010).

\bibitem{Gerritsma11} R. Gerritsma, B. P. Lanyon, G. Kirchmair, F. Z\"ahringer, C. Hempel, J. Casanova, J. J. Garc\'ia-Ripoll, E. Solano, R. Blatt, and C. F. Roos, Phys. Rev. Lett.~\textbf{106}, 060503 (2011).

\bibitem{Longhi}  S. Longhi, Phys. Rev. B \textbf{81}, 075102 (2010).

\bibitem{Dreisow} F. Dreisow, M. Heinrich, R. Keil, A. T\"unnermann, S. Nolte, S. Longhi, and A. Szameit, Phys. Rev. Lett.~\textbf{105}, 143902 (2010).

\bibitem{Keil} R. Keil, J. M. Zeuner, F. Dreisow, M Heinrich, A. T\"unnermann, S. Nolte, and A. Szameit, Nature Comm.~\textbf{4}, 1368 (2012).

\bibitem{Juzeliunas}  G. Juzeli\={u}nas, J. Ruseckas, M. Lindberg, L. Santos, and P. \"Ohberg, Phys. Rev. A \textbf{77}, 011802(R) (2008).

\bibitem{Otterbach} J. Otterbach, R. G. Unanyan, and M. Fleischhauer, Phys. Rev. Lett.~\textbf{102}, 063602 (2009).

\bibitem{Unanyan} R. G. Unanyan, J. Otterbach, M. Fleischhauer, R. Ruseckas, V. Kudria\u{s}ov, and G. Juzeli\={u}nas, Phys. Rev. Lett.~\textbf{105}, 173603 (2010).

\bibitem{Ruseckas} J. Ruseckas, V. Kudria\u{s}ov, G. Juzeli\={u}nas, R. G. Unanyan, J. Otterbach, M. Fleischhauer, Phys. Rev. A \textbf{83}, 063811 (2011).

\bibitem{Casanova1}  J. Casanova, C. Sab\'in, J. Le\'on, I. L. Egusquiza, R. Gerritsma, C. F. Roos, J. J. Garc\'ia-Ripoll, and E. Solano, Phys. Rev. X \textbf{1}, 021018 (2011).

\bibitem{Noh}  C. Noh, B. M. Rodr\'iguez-Lara, and D. G. Angelakis, Phys. Rev. A \textbf{87}, 040102(R) (2013).

\bibitem{Noh1}  C. Noh, B. M. Rodr\'iguez-Lara, and D. G. Angelakis, New J. Phys. \textbf{14}, 033028 (2012).

\bibitem{Jackiw} R. Jackiw and C. Rebbi, Phys, Rev. D \textbf{13}, 3398 (1976).

\bibitem{Su} W. P. Su, J. R. Shrieffer, and A. J. Heeger, Phys, Rev. B \textbf{22}, 2099 (1980).

\bibitem{Niemi} A. J. Niemi and G. W. Semenoff, Phys. Rep.~\textbf{135}, 99 (1986).

\bibitem{TIreview1} M. Z. Hasan and C. L. Kane, Rev. Mod. Phys.~\textbf{82}, 3045 (2010).

\bibitem{TIreview2} X. L. Qi and S. C. Zhang, Rev. Mod. Phys.~\textbf{83}, 1057 (2011).

\bibitem{RuostekoskiJavanainen2002} J. Ruostekoski, G. V. Dunne, and J. Javanainen, Phys. Rev. Lett.~\textbf{88}, 180401 (2002).

\bibitem{JavanainenRuostekoski2003} J. Javanainen and J. Ruostekoski, Phys. Rev. Lett.~\textbf{91}, 150404 (2003).

\bibitem{YefsahZwierlein2013}
T. Yefsah, A. T. Sommer, M. J. H. Ku, L. W. Cheuk, W. Ji, W. S. Bakr, and M. W. Zwierlein, Nature \textbf{499}, 426 (2013).

\bibitem{KitagawaWhite2012}
T. Kitagawa, M. A. Broome, A. Fedrizzi, M. S. Rudner, E. Berg, I. Kassal, A. Aspuru-Guzik, E. Demler, and A. G. White, Nat. Comm.~\textbf{3}, 882 (2012).

\bibitem{YouMundry2007}
C.-Y. Hou, C. Chamon, and C. Mudry, Phys. Rev. Lett. ~\textbf{98}, 186809 (2007);
B. Seradjeh and M. Franz. Phys. Rev. Lett. ~\textbf{101}, 146401 (2008).

\bibitem{RomanovskyLandman2013}
I. Romanonsky, C. Yannouleas, and U. Landman, Phys. Rev. B \textbf{87}, 165431 (2013).

\bibitem{Ryu}  S. Ryu, A. P. Schnyder, A. Furusaki, and A. W. W. Ludwig, New J. Phys. \textbf{12}, 065010 (2010).


\bibitem{Callias} C. Callias, Commun. math. Phys.~\textbf{62}, 213 (1978).

\bibitem{QHE1}  {\it The Quantum Hall Effect}, 
ed.~R. E. Prange and S. M. Girvin (Springer-Verlag New York, 1990).

\bibitem{QHE2}{\it Perspectives in Quantum Hall Effects}, ed.~S. Das Sarma and A. Pinczuk (John Wiley \& Sons, Inc., 1997) 


\bibitem{Nayak} K. P. Nayak, P. N. Melentiev, M. Morinaga, Fam Le Kien, V. I. Balykin, and K. Hakuta,  Opt. Express \textbf{15}, 5431 (2007).

\bibitem{Vetsch10} E. Vetsch, D. Reitz, G. Sague, R. Schmidt, S. T. Dawkins, and
A. Rauschenbeutel, Phys. Rev. Lett.~\textbf{104} 203603 (2010).

\bibitem{HCE1} S. Ghosh, J. E. Sharping, D. G. Ouzounov and A.L. Gaeta,  Phys. Rev. Lett. \textbf{94}, 093902 (2005).

\bibitem{HCE2} T. Takekoshi and R. J. Knize,  Phys. Rev. Lett. \textbf{98}, 210404 (2007).

\bibitem{HCE3} C. A. Christensen, S. Will, M. Saba, G. B. Jo, Y. I. Shin, W. Ketterle, D. Pritchard,  Phys. Rev. A \textbf{78}, 033429 (2008).

\bibitem{HCE4} S. Vorrath, S. A. M\"{o}ller, P. Windpassinger, K. Bongs and K. Sengstock,  New J. Phys. \textbf{12}, 123015 (2010).

\bibitem{Bajcsy11} M. Bajcsy, S. Hofferberth, T. Peyronel, V. Balic, Q. Liang, A. S. Zibrov, V. Vuletic, and M. D. Lukin, Phys. Rev. A \textbf{83}, 063bose830 (2011).


\bibitem{Petrosyan} E. Shahmoon, G. Kurizki, M. Fleischhauer, and D. Petrosyan, Phys. Rev. A \textbf{83}, 033806 (2011).

\bibitem{Hafezi} M. Hafezi, D. E. Chang, V. Gritsev, E. A. Demler and M. D. Lukin, Phys. Rev. A \textbf{85}, 013822 (2012).

\bibitem{Das} P. Das, C. Noh, and D. G. Angelakis, Europhys. Lett.~\textbf{103} 34001 (2013).

\bibitem{Critchfield} C. L. Critchfield, Phys. Rev. D \textbf{12}, 923 (1975).



\bibitem{Fishbane} P. M. Fishbane, S. G. Gasiorowicz, D. C. Johannsen, and P. Kaus, Phys. Rev. D \textbf{27}, 2433 (1983).

\bibitem{Chodos} A. Chodos, R. L. Jaffe, K. Johnson, C. B. Thorn, and V. Weisskopf, Phys. Rev. D \textbf{9}, 3741 (1974).

%
%
%
%
%
  
\end{thebibliography}
\end{document}